\documentclass[useAMS,usenatbib,fleqn]{mn2e}

\usepackage{times}

\usepackage{graphicx}
\usepackage{epsfig}
\usepackage{amssymb,amsmath}
\usepackage{aas_macros}
\usepackage{fixltx2e} 
\title[Crescent black hole images]{A geometric crescent model for black hole images}
\author[Bin Kamruddin \& Dexter]{Ayman Bin Kamruddin$^{1}$\thanks{E-mail: ayman@berkeley.edu} and Jason Dexter$^{1}$\thanks{E-mail:
    jdexter@berkeley.edu}\\
$^{1}$Departments of Physics and Astronomy,
University of California, Berkeley, CA 94720, USA}
\begin{document}

\pagerange{\pageref{firstpage}--\pageref{lastpage}} \pubyear{2012}
\maketitle

\label{firstpage}

\begin{abstract}
The Event Horizon Telescope (EHT), a global very long baseline interferometry array operating at millimetre wavelengths, is spatially resolving the immediate environments of black holes for the first time. The current observations of the Galactic center black hole, Sagittarius A* (Sgr A*), and M87 have been interpreted in terms of either geometric models (e.g., a symmetric Gaussian) or detailed calculations of the appearance of black hole accretion flows. The former are not physically motivated, while the latter are subject to large systematic uncertainties. Motivated by the dominant relativistic effects of Doppler beaming and gravitational lensing in many calculations, we propose a geometric crescent model for black hole images. We show that this simple model provides an excellent statistical description of the existing EHT data of Sgr A* and M87, superior to other geometric models for Sgr A*. It also qualitatively matches physically predicted models, bridging accretion theory and observation. Based on our results, we make predictions for the detectability of the black hole shadow, a signature of strong gravity, in future observations.
\end{abstract}


\section{Introduction}

The Galactic center black hole (Sgr A*) and the supermassive black hole in the center of the Virgo cluster (M87), are the two largest black holes on the sky, with putative event horizons subtending tens of microarcseconds ($\mu$as). Recently, very long baseline interferometry observations at millimetre wavelengths (mm-VLBI) have detected event horizon scale structure in both sources \citep{doeleman2008,doelemanetal2012}. Future observations with this Event Horizon Telescope (EHT) may detect the black hole shadow, the projection of the circular photon orbit at infinity, providing the first direct evidence for an event horizon in the Universe \citep{bardeen1973,falcke}. 

Due to the small number of baselines in the current array, it is not yet possible to create images from the mm-VLBI observations. Instead, models must be fit to the data in the Fourier domain (uv-plane). Geometric models like Gaussian brightness distributions and constant intensity annuli (rings) have been fit to the data \citep{doeleman2008,fishetal2011,brodericketal2011,doelemanetal2012} as well as ray traced images from theoretical accretion flow and jet models  \citep{broderick2009,broderickloeb2009,brodericketal2011,moscibrodzka2009,dexter2009,dexteretal2010,dexteretal2012,dexterfragile2012,shcherbakovetal2012}. The former are physically unmotivated, while the latter are subject to significant systematic uncertainties. These uncertainties include a poor understanding of the distribution function of the radiating electrons, the circularization radius and infall rate of gas from large scales \citep{cuadraetal2006,pangetal2011}, the orientation of the inner disc relative to the black hole \citep{fragile2007}, and the dynamical importance of magnetic fields \citep{mckinneyetal2012}. 

Despite these uncertainties, the emission in models which fit current data arises from very close to the black hole \citep[$r < 10$ M][]{moscibrodzka2009}, where the relativistic effects of Doppler beaming from orbital motion and light bending from strong gravity tend to dominate the black hole images. As a result, many different theoretical models lead to images which are ``crescent'' shaped: Doppler beaming causes emission from approaching material to be brighter than that of receding material, while light bending causes the back of the accretion disc or torus to appear above and below the black hole \citep{bromley2001,noble2007,moscibrodzka2009,huang2009,yuan2009,broderick2009,brodericketal2011,dexteretal2010,dexteretal2012,dexterfragile2012,straubetal2012}.

Motivated by these results, we present a simple geometric model for crescent black hole images (\S\ref{sec:methods}). We fit the model to existing data of Sgr A* and M87, constrain its parameters, and show that it statistically outperforms geometric models used previously (\S\ref{sec:results}). In \S\ref{sec:discussion}, we discuss the implications for constraining the model with future observations, and show that the crescent model does a reasonable job reproducing a range of accretion flow and jet images from numerical simulations. We also show that if the model is correct, the black hole shadows in Sgr A* and M87 may be accessible to observations in the near future.

\begin{figure*}
\begin{center}
\begin{tabular}{lll}
\includegraphics[width=2in]{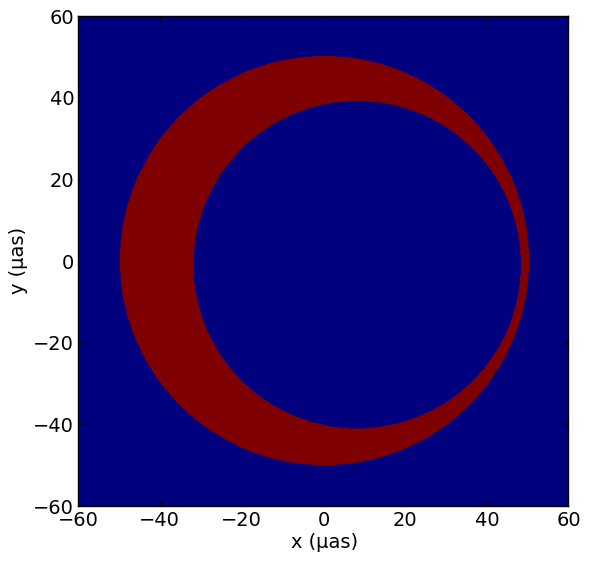}&
\includegraphics[width=2in]{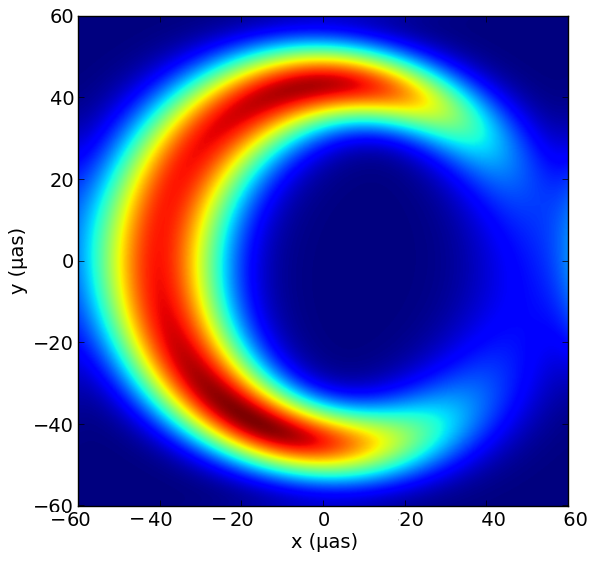}&
\includegraphics[width=2.4in]{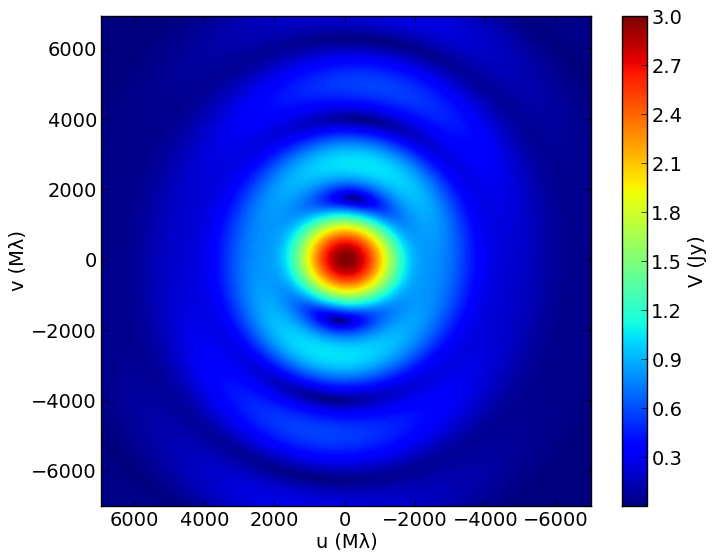}
\end{tabular}
\end{center}
\caption{\label{demo_crescent}Sample crescent image (left), blurred image (center), and blurred visibility amplitude (right). The crescent parameters are $R_p = 50 \mu$as, $R_n = 40 \mu$as, $a = 8 \mu$as, $b = 1 \mu$as, $V_0 = 3$ Jy.}
\end{figure*}

\section{Crescent Model}
\label{sec:methods}

The crescent model is created by subtracting out a disc from the inside of a larger disc, both of constant intensity. Only discs where the smaller disc lies inside the larger disc have been considered, to allow the Fourier transform to be evaluated analytically.  The larger disc is centered at the origin. The five parameters used to describe the model are the radius of the larger (smaller) disc, $R_p$ ($R_n$); the $x$ and $y$ positions of the centroid of the inner disc, $a$ and $b$; and the total flux, $V_0$. All positions are measured in microarcseconds ($\mu$as), while the total flux is measured in Jy.

The requirement of real and positive image intensity leads to constraints on the parameters: $V_0$, $R_p$, $R_n > 0$, and $R_p - R_n - \sqrt{a^2+b^2} > 0$, where the last constraint ensures that the smaller disc lies completely within the larger one. For a given set of parameters, we express the spatial intensity of the crescent as a function of spatial coordinates, $I(x,y)$. The Fourier transform (FT) of the crescent, defined as

\begin{equation}
V(u,v) = \int \int dx dy I(x,y) e^{-2\pi i (ux+vy)/\lambda},
\end{equation}

\noindent where $V(u,v)$ is the complex visibility as a function of spatial frequencies $u$ and $v$ in units of the observed wavelength $\lambda$, can be calculated analytically using the FT of a single disc of radius $R$, $V_d (u,v)$:

\begin{equation}
  V_d (u,v; I_0, R) = \pi R^2 I_0 \frac{2 J_1 (k R)}{k R},
\end{equation}

\noindent where $k \equiv 2 \pi \sqrt{u^2+v^2} / \lambda$, $J_1 (x)$ is the Bessel function of the first kind, and $I_0$ is the constant surface brightness of the disc. The crescent model is the difference of two discs, one displaced from the origin:  

\begin{align}
 &V_c(u,v;\hspace{2pt} V_0, R_p, R_n, a, b) \nonumber\\
&= V_d (u, v; I_0, R_p) - e^{-2\pi i (a u + b v) / \lambda} V_d (u, v; I_0, R_n)\nonumber\\
&= 2 \pi I_0 \left[\frac{R_p J_1 (k R_p)}{k} - e^{-2\pi i (a u + b v) / \lambda} \frac{R_n J_1 (k R_n)}{k}\right]\nonumber\\
&= \frac{2 V_0}{k (R_p^2 - R_n^2)} \left[R_p J_1 (k R_p) - e^{-2\pi i (a u + b v) / \lambda} R_n J_1 (k R_n)\right],
\end{align}

\noindent where in the last step the constant intensity $I_0$ is replaced with the total flux of the crescent, $V_0 = \pi (R_p^2 - R_n^2) I_0$.

\begin{figure*}
\begin{center}
\begin{tabular}{ll}
\includegraphics[width=3in]{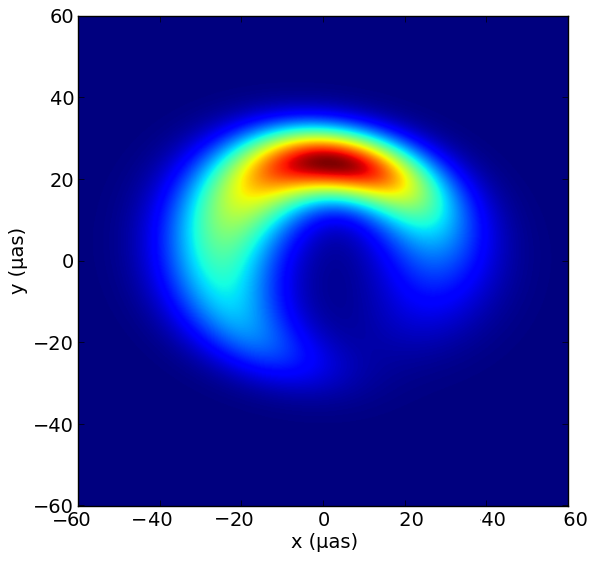}&
\includegraphics[width=3in]{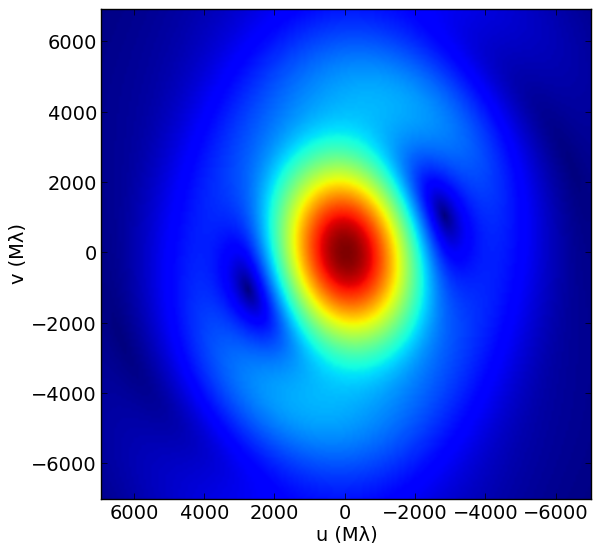}
\end{tabular}
\end{center}
\caption{\label{best_crescent}Blurred best fitting crescent image to all of the Sgr A* data (left) and the amplitude of its Fourier transform (right).}
\end{figure*}

\begin{figure}
\includegraphics[width=3in]{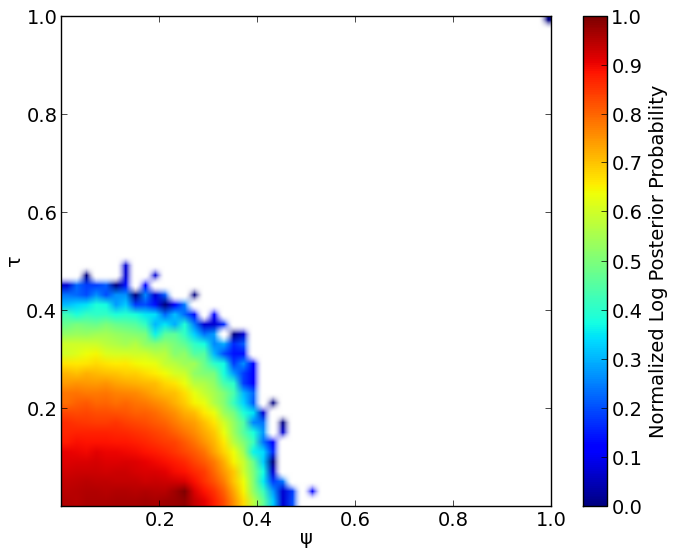}
\caption{\label{contourplot}Normalized log posterior probability vs. $\tau$ and $\psi$ for fits of the crescent model to all of the Sgr A* mm-VLBI data. The favored models are highly asymmetric ($\tau \ll 1$, unlike a ring) and have large subtracted regions ($\psi \ll 1$, unlike a disc).}
\end{figure}

\begin{figure}
\begin{center}
\begin{tabular}{lll}
\includegraphics[width=1.5in]{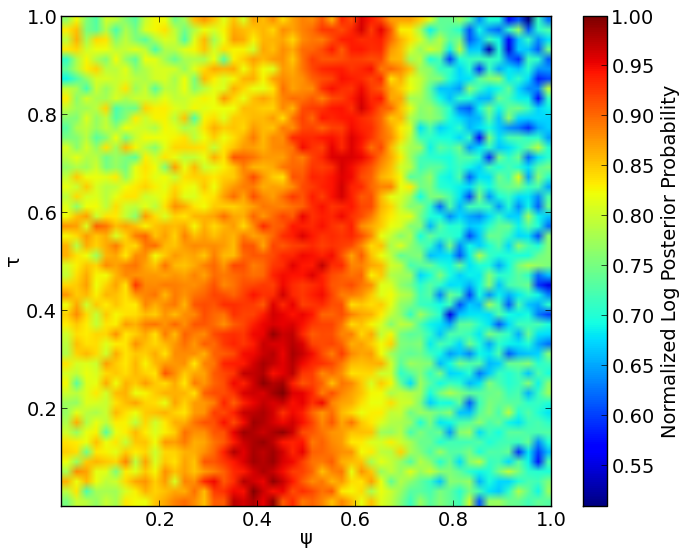}&
\includegraphics[width=1.5in]{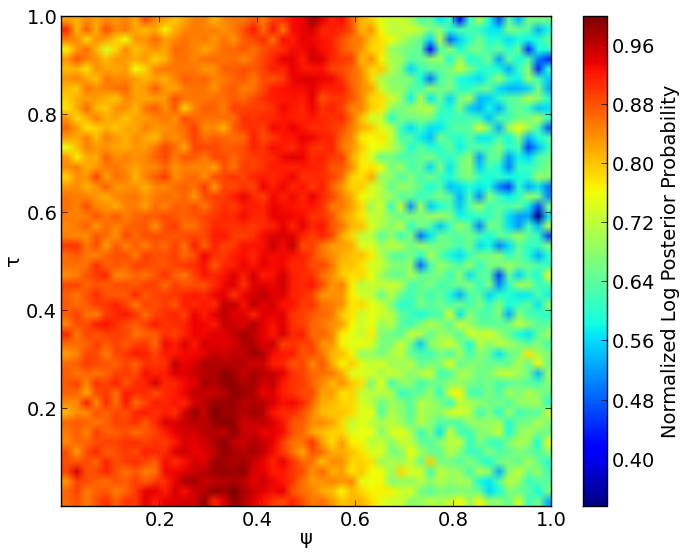}\\
\includegraphics[width=1.5in]{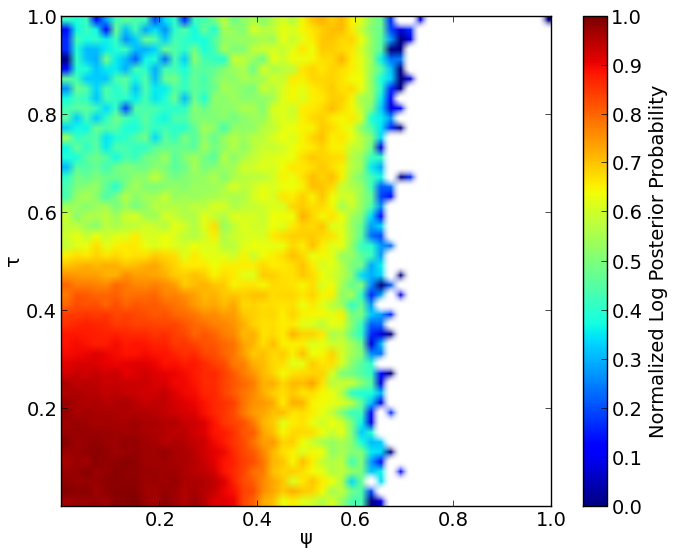}&
\includegraphics[width=1.5in]{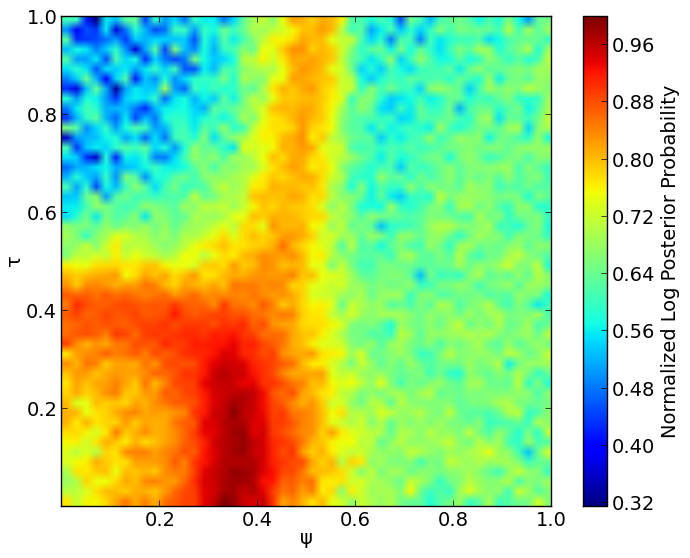}
\end{tabular}
\end{center}
\caption{\label{eachdaycontours}Normalized log posterior probability vs. $\tau$ and $\psi$ for fits of the crescent model to each of the four different Sgr A* days of mm-VLBI data: 2007 (top left) and three days in 2009. The allowed ranges overlap between days, and there is no evidence for structural variability between days despite changes in total flux during the 2009 campaign.}
\end{figure} 

\begin{figure*}
\begin{center}
\begin{tabular}{ll}
\includegraphics[width=2.35in]{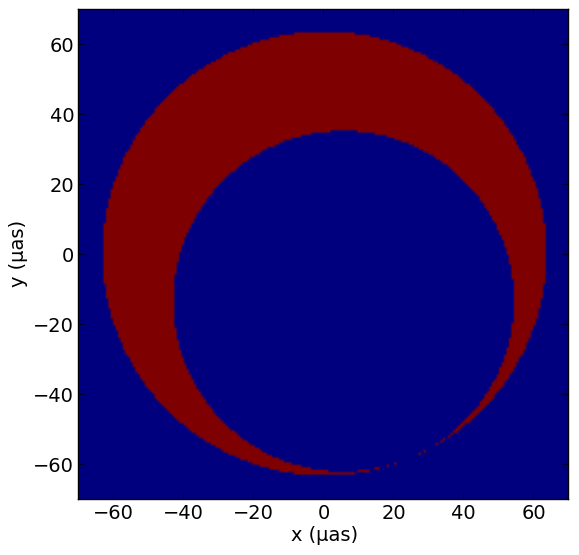}&
\includegraphics[width=3in]{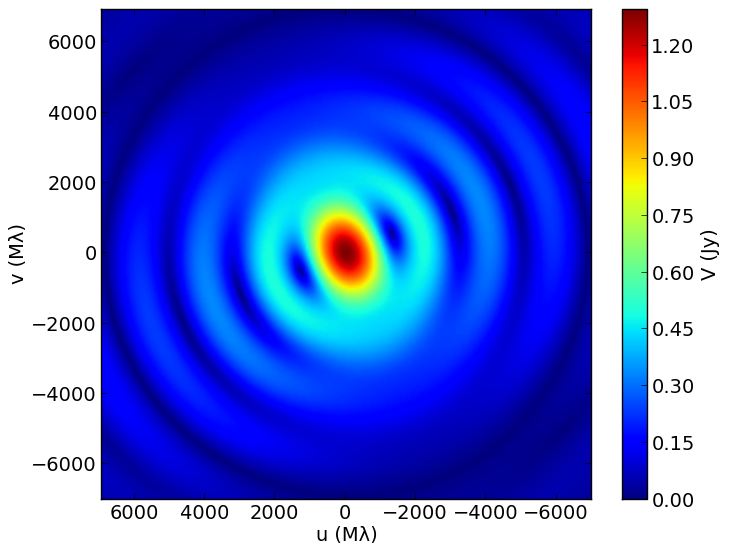}
\end{tabular}
\end{center}
\caption{\label{best_m87}Best fitting crescent image to all of the M87 data (left) and the amplitude of its Fourier transform (right). The crescent is constant intensity since in this case there is no blurring from interstellar scattering. Note that the overall size of the best fitting crescent is larger in this case than for Sgr A*.}
\end{figure*}

\begin{figure}
\includegraphics[width=3in]{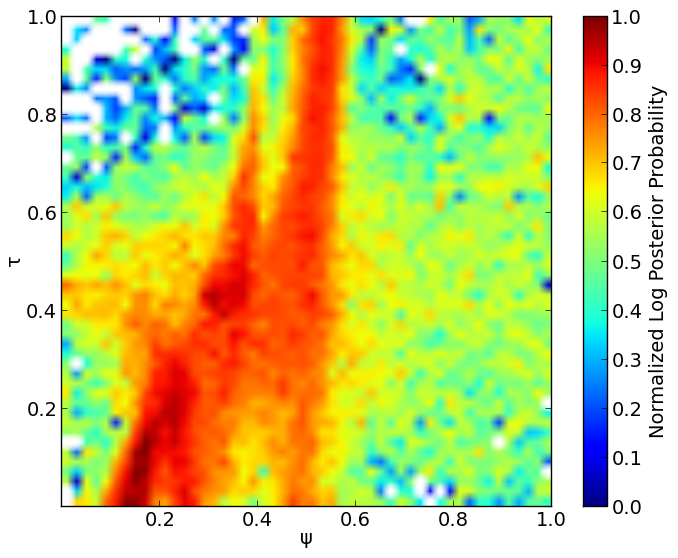}
\caption{\label{M87_contourplot}Normalized log posterior probability vs. $\tau$ and $\psi$ for fits of the crescent model to all of the M87 mm-VLBI data. The constraints are much weaker than for the Sgr A* data (FIgure \ref{contourplot}), although disc-like models ($\psi = 1$) are still ruled out.}
\end{figure}

Interstellar scattering from free electrons effectively blurs images of Sgr A*. This effect is taken into account by convolving the images (multiplying the visibilities) by an asymmetric Gaussian \citep{bower2006,fish2009}. Performing the inverse FT then gives the blurred crescent image. A sample crescent image, blurred image, and visibility are shown in Figure \ref{demo_crescent}. The sample crescent's visibility amplitude decreases monotonically with baseline length on left-to-right orientations, with the ringing present from Bessel functions when looking top-to-bottom. There is a pronounced minimum which corresponds to the size of the gap in the middle. In many theoretical images, this gap corresponds to the black hole shadow, the effective cross section of the black hole to photons  \citep{bardeen1973}.

An equivalent set of parameters is more convenient for fitting the crescent model to data: 

\begin{align}\label{params}
R &\equiv R_p, \hspace{6pt} \psi \equiv 1 - R_n / R_p, \nonumber\\\tau &\equiv 1 - \frac{\sqrt{a^2 + b^2}}{R_p - R_n}, \hspace{6pt} \phi \equiv \tan^{-1} \frac{b}{a}.
\end{align}

\noindent and the normalization, $V_0$, is unchanged. The new parameters describe the overall size ($R$), relative thickness ($\psi$), degree of symmetry ($\tau$), and orientation ($\phi$) of the crescent. The parameter constraints then simply become $V_0, R > 0$, $\tau > 0$, $\psi < 1$, $-\pi < \phi < \pi$. These simplified constraints allow for convergence in the posterior probability distributions with less computation time due to a higher acceptance rate.

For comparison, we also use previously considered geometric models. The ``ring'' model is a subset of crescent models with $a=b=0$ or $\tau=1$. \citet{brodericketal2011} also compared radiatively inefficient accretion flow models \citep[RIAFs,][]{yuanquataert2003} to an asymmetric Gaussian, whose visibility is:

\begin{equation}
  V_g (u,v) = V_0 e^{-a u^2 + 2 b u v - c v^2}
\end{equation}

\noindent  where

\begin{align}
  a &=\frac{\cos^2{\theta}}{2 \sigma_u^2} + \frac{\sin^2{\theta}}{2 \sigma_v^2}\\
  b &=-\frac{\sin{2 \theta}}{4 \sigma_u^2} + \frac{\sin{2 \theta}}{4 \sigma_v^2}\\
  c &=\frac{\sin^2{\theta}}{2 \sigma_u^2} + \frac{\cos^2{\theta}}{2 \sigma_v^2}
\end{align}
\noindent  and 
\begin{align}
  \sigma_{u,v} & = \frac{1}{2 \pi \sigma_{x,y}}\\
\end{align}

\noindent where $\sigma_{x,y}$ are the widths along the axes, which are rotated counterclockwise from the x-axis by an angle $\theta$, and $V_0$ is the total flux.

Following \citet{brodericketal2011}, we redefined the parameters as:

\begin{align}
 \sigma^2 &= \frac{\sigma_x^{-2} + \sigma_y^{-2}}{2}\\
 A &=  \frac{\sigma^2}{2} (\sigma_x^{-2} - \sigma_y^{-2})
\end{align}
 
\noindent where $\sigma$ gives the overall size and $A$ indicates the degree of asymmetry. We verified that our model fitting results to the mm-VLBI data for the  Gaussian reproduce those of \citet{brodericketal2011}. The symmetric Gaussian, used to measure the source sizes of both Sgr A* \citep{doeleman2008,fishetal2011} and M87 \citep{doelemanetal2012}, is a subset with $\sigma_x = \sigma_y$ and fixed $\theta$ ($A=0$). This model is strongly disfavored compared to the asymmetric Gaussian \citep{brodericketal2011} and we do not consider it separately.

\begin{figure}
\includegraphics[width=3.5in]{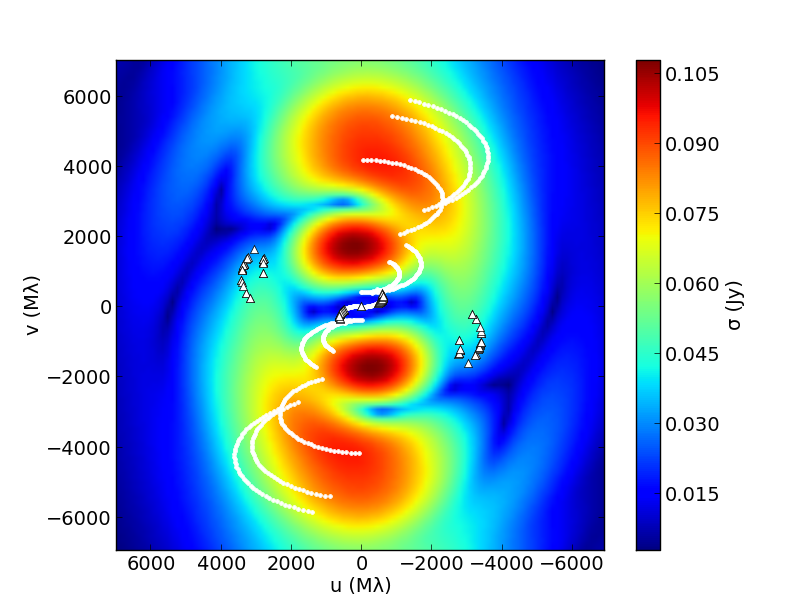}
\caption{\label{stddev}Variance of possible crescent images of Sgr A* in the uv-plane, weighted by their posterior probability. Regions of large variance correspond to the most promising locations for constraining the crescent model with future mm-VLBI observations. The locations corresponding to current and future observations are overplotted as triangles and solid lines.}
\end{figure}

\begin{figure}
\includegraphics[width=3in]{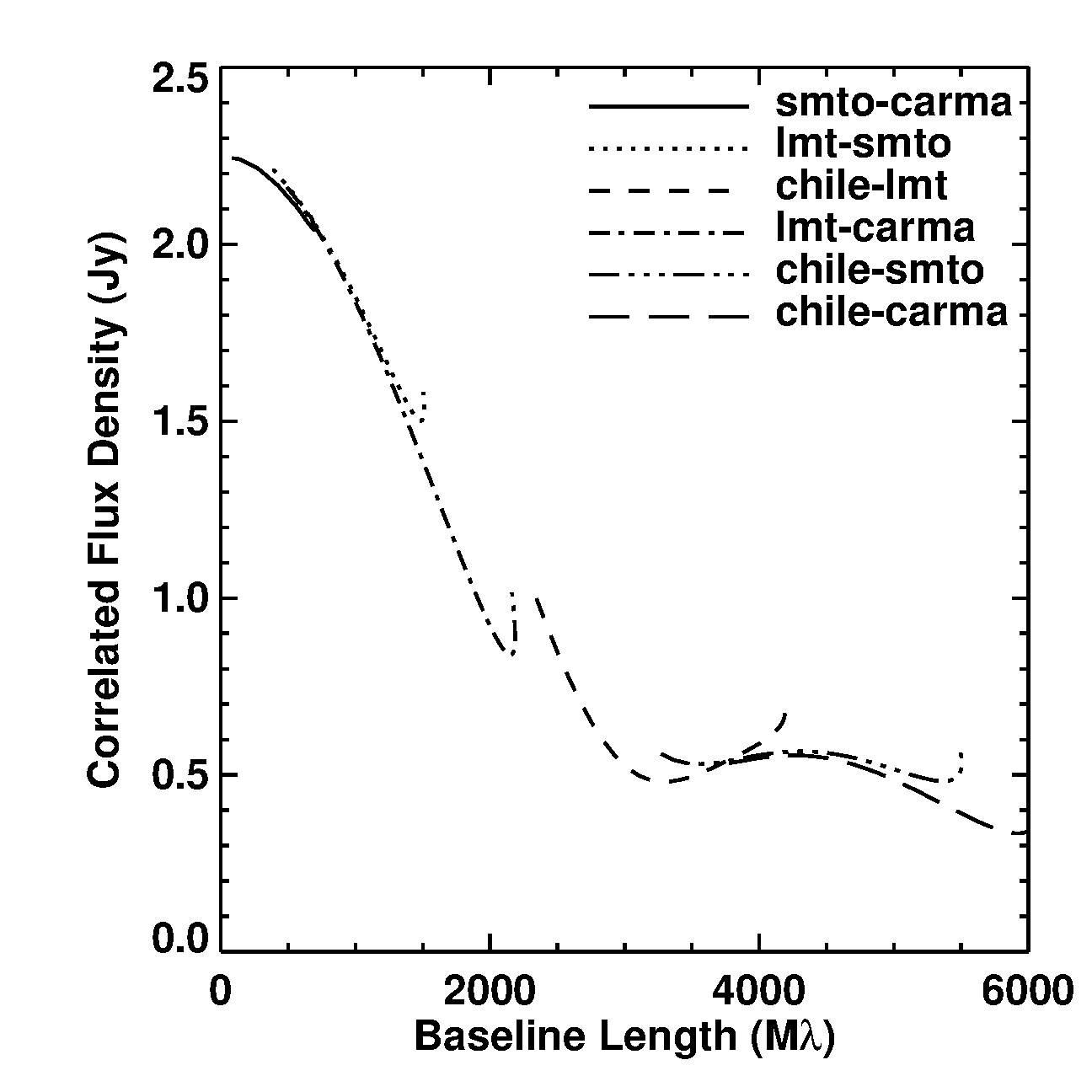}\\
\includegraphics[width=3in]{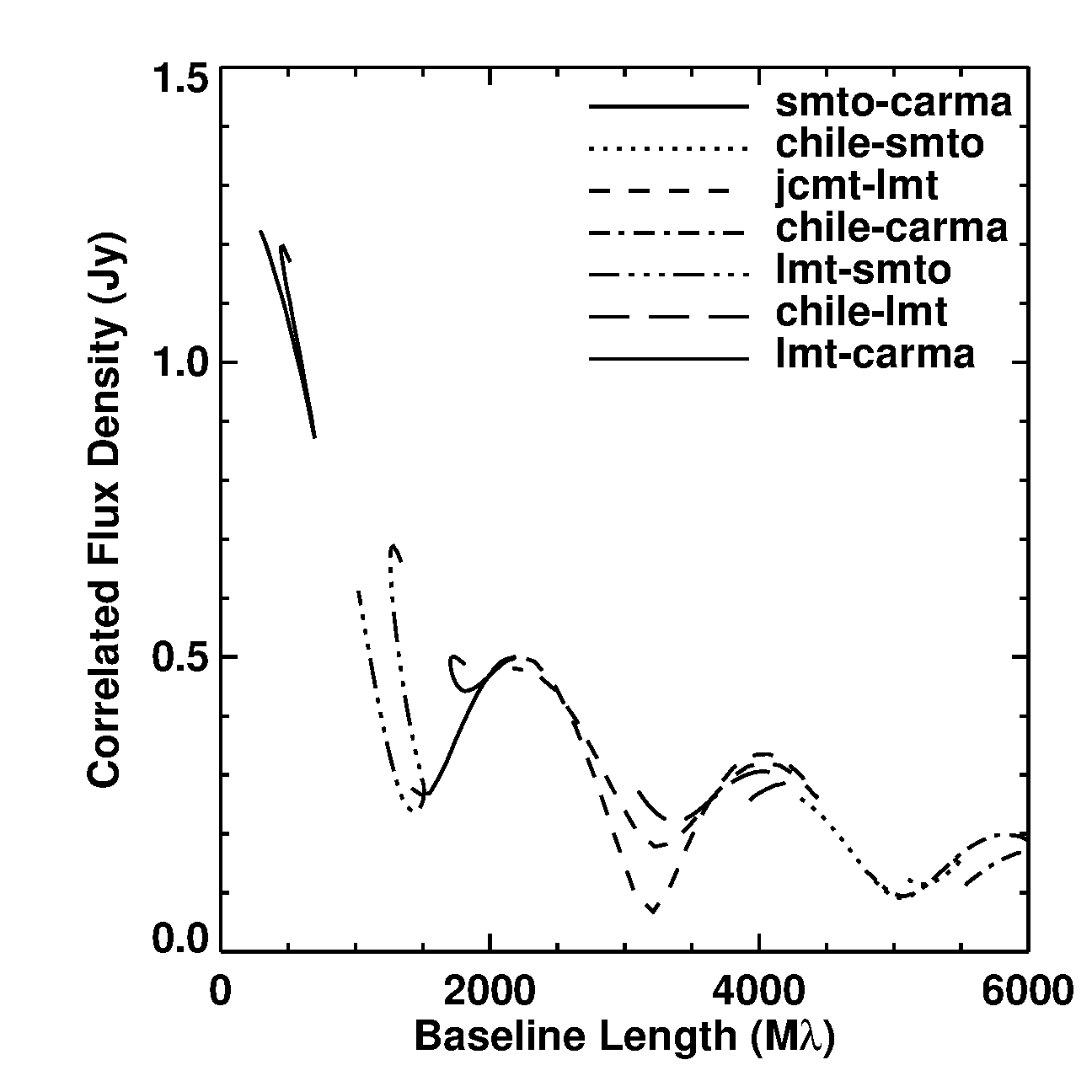}
\caption{\label{interp}Predicted visibility amplitude vs. baseline length for the best fitting crescent models for Sgr A* (top) and M87 (bottom) on future baselines including telescopes in Chile (ALMA/APEX) and Mexico (LMT) as well as existing telescopes in Arizona (SMTO) and California (CARMA). The black hole shadow produces the distinct minima in the M87 curves.}
\end{figure}

\begin{figure*}
\begin{center}
\begin{tabular}{lll}
\includegraphics[width=2in]{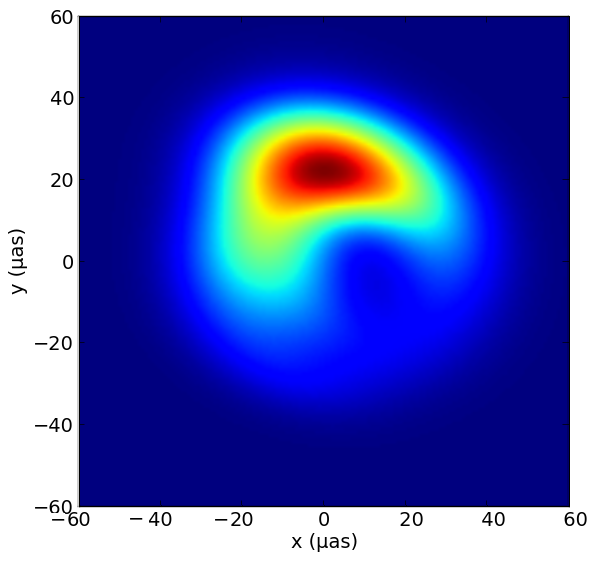}&
\includegraphics[width=2in]{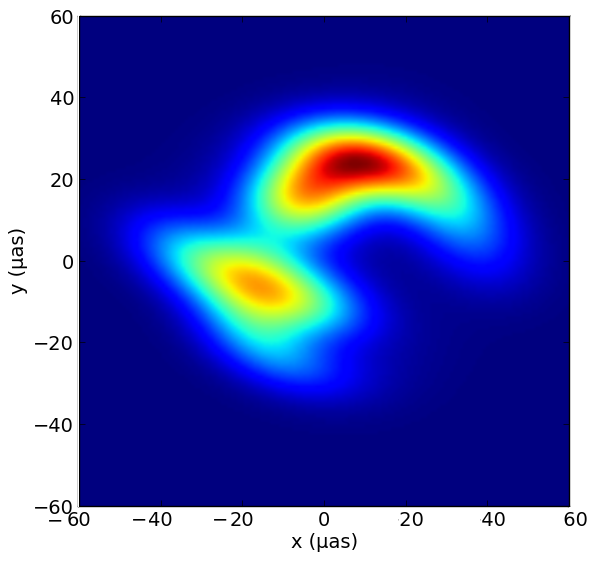}&
\includegraphics[width=2in]{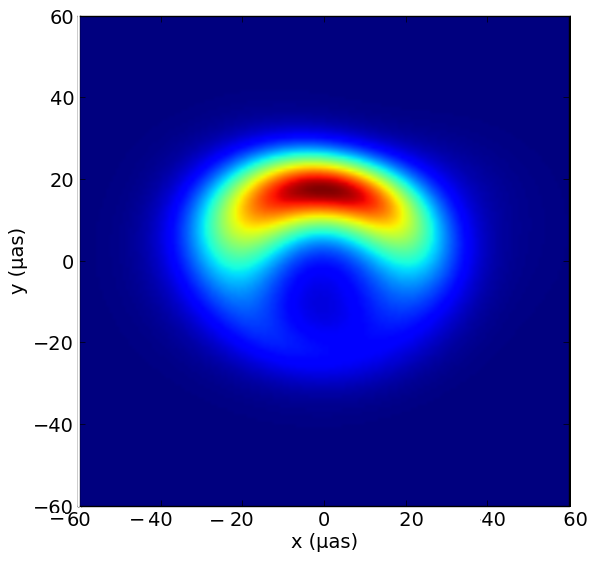}\\
\includegraphics[width=2in]{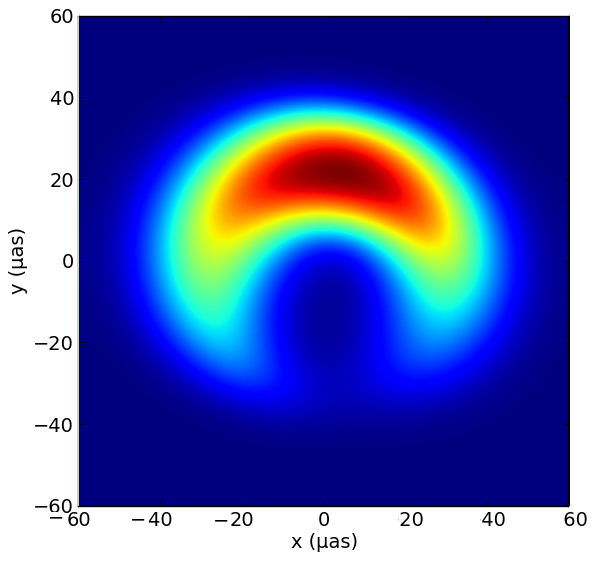}&
\includegraphics[width=2in]{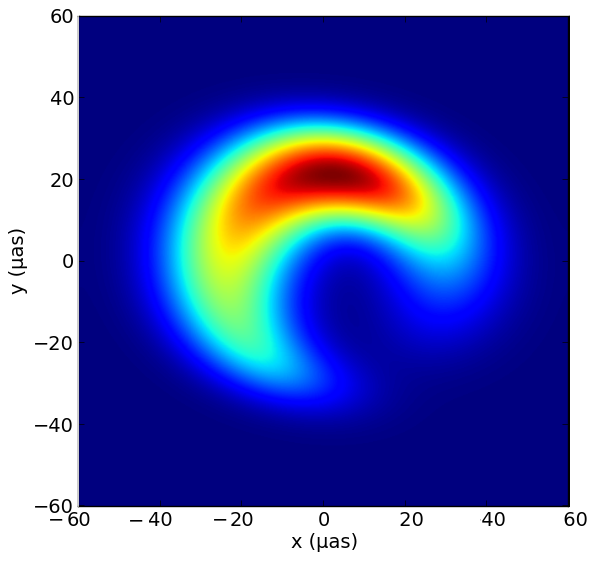}&
\includegraphics[width=2in]{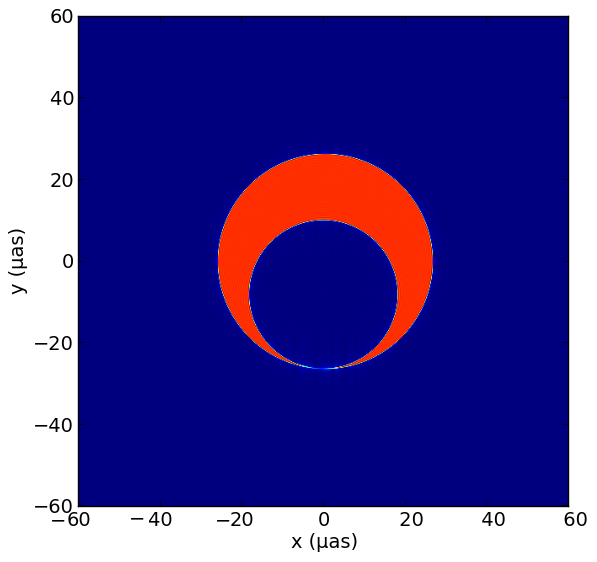}\\
\end{tabular}
\end{center}
\caption{\label{simulations}Crescent model fits (bottom) to simulated images of black hole accretion flows and jets (top). The simulated images are models MBD from \citet{dexteretal2010} (left) and 915h from \citet{dexterfragile2012} (center) of Sgr A*, and J2 from \citet{dexteretal2012} of M87 (right). The left two Sgr A* models as well as their crescent fits have been blurred to account for the effects of interstellar scattering. In all cases, the crescent model fits capture the main orientation and size of the simulated image.}
\end{figure*} 

\section{Model Fitting}

We fit the above models to the mm-VLBI data for Sgr A* and M87. The data consist of visibility amplitudes measured at many ($u$, $v$) locations. The Sgr A* data were taken on 4 separate days from two campaigns \citep{doeleman2008,fishetal2011}. The total flux in the Sgr A* data ($V(u\simeq0,v\simeq0)$) varied by $\simeq 30\%$ between days in the \citet{fishetal2011} observations, while there is no evidence for differences in total flux between days in the observations of M87. This is consistent with the relevant physical timescales associated with the two black hole masses: minutes to hours \citep[Sgr A*, $M_{\rm bh} \simeq 4\times10^{6} M_\odot$, e.g.,][]{gillessen2009} and days to years \citep[M87, $M_{\rm bh} \simeq 6\times10^{9} M_\odot$,][]{gebhardtetal2011}

We calculate the posterior probability distribution over all parameters of each geometric model. The distributions are sampled using \textsc{emcee}, a public implementation of the \citet{goodmanweare} Affine Invariant Markov Chain Monte Carlo (MCMC) Ensemble sampler \citep{emcee}. This enables us to determine the best fitting revised parameters (Eqs. \ref{params}) of the crescent. The $\chi^2$  of a model fitting the data is given by:

\begin{equation}
\chi^2 = \sum_i \frac{(V_{\rm model, i}- V_{\rm obs, i})^2}{\Delta V_{\rm obs, i}^2},
\end{equation}

\noindent where the sum is over all observed $V(u,v)$. We assume uniform priors on all parameters for all models, so that the posterior probability is simply proportional to $\chi^2$. 

The \textsc{emcee} algorithm uses many separate ``walkers'' to sample the parameter space, each using an MCMC algorithm with a set number of trials. The total number of samples is then given by the product of the number of walkers with the number of trials per walker. We initialized the MCMC runs by doing a "burn-in" starting from a guess for the crescent using 1000 walkers with 100 trials each, and then used the final points from that run to seed a larger one. The results were well converged results in all cases using 1000 walkers with 1000 trials each, with respect to decreasing/increasing the number of walkers/trials, and changing the starting location in the parameter space. The resulting probability distributions also appear relatively smooth (see below), a sign of convergence.

First we fit to all the data from Sgr A* at once to get the overall parameter ranges. The distributions are well converged, and the allowed parameter ranges are given in Table \ref{sgraranges}. The best fitting crescent image of Sgr A*, including the blurring from interstellar scattering, is shown in Figure \ref{best_crescent} along with the amplitude of its Fourier transform. The best fitting crescent does not resemble a simpler ring or disc model. This can also be seen from a 2D contour plot of the log probability vs. $\tau$ and $\psi$ (Figure \ref{contourplot}). The probability density is concentrated near where both parameters are small, which points the high preference of crescent-like structures over ring-like or disc-like ones.  The preference for $\tau$ and $\psi$ values close to zero suggests that there may be better crescents with negative $\tau$, which are crescents where the inner disc does not completely lie within the outer disc (models where the subtracted region is not circular). 

The reduced chi-squared from fitting all data simultaneously is relatively high ($\gtrsim 2$), and this is true for all models. This is because the total flux of Sgr A* varied significantly between different days during the second mm-VLBI campaign \citep{fishetal2011}. To find a satisfactory fit, one or more parameters should then be allowed to vary between days. We test for structural changes between days by comparing allowed crescent parameter ranges fitting the data from each day separately in the same way as \citet{brodericketal2011}. The resulting 2D contour plots of probability vs. $\tau$ and $\psi$ are shown in Figure \ref{eachdaycontours}. The $68\%$ confidence intervals overlap, so that there is no strong evidence for structural variations between days. We then fix the structure to the best fit parameters from using all the data, and only let $V_0$ vary between days to find better fits. This procedure is repeated for the Gaussian and ring models, similarly finding no strong evidence for structural variations between days. The resulting $\chi^2$ values for the best fitting models are listed in Table \ref{sgrafit}. The number of parameters k represents the total number of parameters for the runs. For the Crescent, for instance, four different values of $V_0$ for the four days give four parameters; and $R$, $\tau$, $\psi$ and $\psi$ are the other four parameters and so the total is 8. The same goes for the Gaussian and Ring.  All geometric models give satisfactory fits to the mm-VLBI data (reduced $\chi^2 \simeq 1$) when allowing the total image intensity to vary between days.

We repeat the same exercise fitting to the M87 data. In this case there is no evidence for strong interstellar scattering, and so we do not blur the model images. There is no evidence for flux or structural variations between days in the M87 data, and we do not expect variability on these timescales since the light crossing time is $\simeq 1$ day. The best fitting crescent to these data are shown in Figure \ref{best_m87}. A contour plot of the log posterior probability vs. $\tau$ and $\psi$ is shown in Figure \ref{M87_contourplot}, and the allowed parameter ranges are listed in Table \ref{m87ranges}. As with Sgr A*, the best-fitting crescent does not resemble a ring or a blob, instead preferring relatively small $\tau$ and $\psi$. The $\chi^2$ values are again listed in Table \ref{m87fit}: all models produce excellent fits to the M87 data.

\begin{table}
\caption{Parameter Ranges for fit to Sgr A* (fit to entire data set) \label{sgraranges}}
\begin{small}
\begin{center}
\begin{tabular}{lccccccc}
        \hline
	\hline
Parameter & Best fit value & $+68\%$ & $+95\%$ & $-68\%$ & $-95\%$\\
        \hline
R & $26.99$ & $6.99$ & $8.93$ & $2.99$ & $4.31$\\
$\tau$ & $.02$ & $.30$ & $.38$ & $.02$ & $.02$\\
$\psi$ & $.05$ & $.30$ & $.35$ & $.05$ & $.05$\\
$\phi$ & $1.21$ & $.14$ & $.19$ & $.39$ & $.64$\\
I & $2.25$ & $.07$ & $.10$ & $.07$ & $.10$\\
	\hline
\end{tabular}
\end{center}
\end{small}
\end{table}

\begin{table}
\caption{Goodness of Fit Summary for Sgr A* (varying $V_0$ between days) \label{sgrafit}}
\begin{small}
\begin{center}
\begin{tabular}{lccccccc}
        \hline
	\hline
Model & k & $\chi^2$ & $\chi^2 / \rm d.o.f.$ & AIC & BIC & $w_{\rm AIC}$ & $w_{\rm BIC}$\\
        \hline
Crescent & $8$ & $49.66$ & $0.80$ & $68.02$ & $83.64$ & $1$ &   $1$\\
Gaussian & $7$ & $63.66$ & $1.01$ & $79.46$ & $93.39$ & $0.0032$ & $0.0076$ \\
Ring     & $6$ & $65.97$    & $1.03$ & $79.31$    & $91.46$   & $.0035$    & $.0200$\\
	\hline
\end{tabular}
\end{center}
\end{small}
\end{table}

\begin{table}
\caption{Parameter Ranges for fit to M87 \label{m87ranges}}
\begin{small}
\begin{center}
\begin{tabular}{lccccccc}
        \hline
	\hline
Parameter & Best fit value & $+68\%$ & $+95\%$ & $-68\%$ & $-95\%$\\
        \hline
R & $63.56$ & $4.07$ & $6.14$ & $7.94$ & $9.43$\\
$\tau$ & $.01$ & $.44$ & $.94$ & $.01$ & $.01$\\
$\psi$ & $.22$ & $.21$ & $.35$ & $.22$ & $.22$\\
$\phi$ & $1.20$ & $.23$ & $1.30$ & $1.85$ & $2.25$\\
I & $1.31$ & $.07$ & $.10$ & $.25$ & $.27$\\
	\hline
\end{tabular}
\end{center}
\end{small}
\end{table}

\begin{table}
\caption{Goodness of Fit Summary for M87 \label{m87fit}}
\begin{small}
\begin{center}
\begin{tabular}{lccccccc}
        \hline
	\hline
Model & k & $\chi^2$ & $\chi^2 / \rm d.o.f.$ & AIC & BIC & $w_{\rm AIC}$ & $w_{\rm BIC}$\\
        \hline
Crescent & $5$ & $55.14$ & $0.56$ & $65.76$ & $78.31$ & $1.0$ & $1.0$\\
Gaussian & $4$ & $58.55$ & $0.59$ & $66.96$ & $77.09$ & $0.55$ & $1.84$ \\
Ring & $3$ & $63.39$ & $0.63$ & $69.63$ & $77.29$ & $0.14$ & $1.67$\\
	\hline
\end{tabular}
\end{center}
\end{small}
\end{table}

\label{sec:results}

To quantitatively compare the fit quality of the different geometric models which are not all nested (Gaussian and crescent) and which have different numbers of free parameters, we follow \citet{brodericketal2011} and use information criteria (AIC/BIC), defined as:

\begin{align}
  AIC&=\chi^2 + 2k + 2k(k+1)/(N-k-1)\\
  BIC&=\chi^2 + k \ln{N}
\end{align}

\noindent where $\chi^2$ is the minimum value found for each model across all parameter values (listed in Tables \ref{sgrafit} and \ref{m87fit}), $N$ is the number of data points, and $k$ is the number of free parameters. Small values of AIC/BIC indicate a good fit, as these numbers are basically $\chi^2$ with an added penalty for free parameters. A difference $\Delta \rm IC > 5$, $10$ is usually taken as strong or definitive evidence in favor of the model with the lower IC value. As in \citet{brodericketal2011}, we also compare the performance of a model with information criteria IC1 relative to a second model with IC2 with the ratio $e^{(IC1-IC2)/2}$, which indicates the probability that model 2 does as good of a job as model 1. Note that to compute this odds ratio we should integrate over all parameter space for each model, but given uncertainties in the priors and parameter space volumes it is not clear this would be a better approach than just comparing the best fits between models.

The odds ratio calculated in this way favors the crescent at $\thicksim 2-3 \sigma$ significance. This is very similar to the result found by \citet{brodericketal2011} using a semi-analytic RIAF model \citep{yuanquataert2003} instead of the crescent. We do the same for the ring, and find a similar odds ratio as for the Gaussian. For M87 the lowest $\chi^2$ value is from the crescent model, but the AIC/BIC do not significantly favor the crescent over the Gaussian or ring models in this case. In all cases, using AIC/BIC is likely a conservative approach: comparing the nested ring and crescent models using F-tests or likelihood ratio tests leads to small ($\lesssim 3 \sigma$) probability values for the ring, even for M87 where the odds ratios show little evidence for the crescent model being superior. This finding is consistent with Figures \ref{contourplot} and \ref{M87_contourplot}, where the probability values are very small for ring images ($\tau=1$). 

\section{Discussion}
\label{sec:discussion}

We have proposed a simple geometric model for the common crescent morphology for black hole images resulting from the combined relativistic effects of Doppler beaming and gravitational light bending. Unlike previously used geometric models (Gaussians and rings), this one is physically motivated, and in addition provides a statistically better description ($\simeq 2.5-3.0\sigma$ significance) of the Sgr A* mm-VLBI data even with limited sensitivity and coverage (Table \ref{sgrafit}). Thus far, the M87 data cannot distinguish between ring, Gaussian, and crescent models (Table \ref{m87fit}).

The results for Sgr A* have interesting implications for the angular momentum content and viewing geometry of the accretion flow or jet base. Recent large scale simulations find that strong magnetic fields can prevent accreting gas from circularizing \citep{pangetal2011}, which would lead to a circularly symmetric image on timescales longer than the orbital time. Since the data disfavor a ring, this is also evidence for angular momentum in the accretion flow. Similarly, regardless of the angular momentum content a face-on configuration leads to a ring-like image. This geometry is already disfavored because face-on configurations have lower optical depths at $1.3$ mm and steeper than observed spectral indices between $0.4$ and $1.3$ mm \citep{marronephd,moscibrodzka2009}, but our results provide additional, independent evidence against a face-on configuration unless the image morphology is non-axisymmetric on times longer than the orbital time \citep[e.g., from disc tilt,][]{dexterfragile2012}. The data also favor the crescent model over a Gaussian, presumably because of the curvature in crescent images. This curvature arises in theoretical images as a result of strong light bending near the black hole, and therefore the statistical superiority of the crescent model could be interpreted as observational evidence for strong gravitational light bending in Sgr A*.

The range of $R$ preferred for the M87 data is significantly larger ($> 3\sigma$) than that previously found for a symmetric Gaussian model by \citet{doelemanetal2012}. They used this inferred size as evidence for a prograde black hole in M87, arguing that the jet size should correspond to the lensed size of the innermost stable circular orbit of the black hole. However, since the data cannot currently distinguish between various models, the inferred size from any given model is not necessarily physically meaningful. For example, combining the parameter $R$ from the best fitting crescent with the black hole mass from \citet{gebhardtetal2011} gives a size of $\simeq 8$ Schwarzschild radii, which would correspond to the ISCO of a black hole with spin $a \simeq -0.6$.

The crescent model can be tested with future mm-VLBI observations, both incorporating phase information and additional telescopes (baselines). Although the models have only been fit to visibility amplitudes, a closure phase of $\pm 40^\circ$ has been reported for Sgr A* on the triangle of current telescopes \citep{fishetal2011}. The best fitting crescent model has closure phases on this triangle of $0-50^\circ$, consistent with the observations. Unpublished observations show closure phases much closer to zero (V. Fish, private communication), and future data including closure phase information will place important additional constraints on the models. 

We can also assess the prospects for constraining the crescent model with visibility amplitude observations on additional baselines. In Figure \ref{stddev}, we show the weighted standard deviation of the model visibility amplitudes as a function of position in the uv-plane. Regions of large standard deviation correspond to locations where the models predict a wide range of amplitudes, and thus where they can be best constrained with future observations. The most promising baselines have lengths similar to or even shorter than those in the current array, but at nearly orthogonal orientations. An example of such a baseline would be between Chile (e.g., ALMA or APEX) and Mexico (LMT). These results are similar to previous maps based on semi-analytic RIAF models \citep{fish2009} and relativistic MHD simulations \citep{dexteretal2010}. 

One major goal of the observations is to detect the black hole shadow, corresponding to the projection of the circular photon orbit on the sky \citep{bardeen1973,falcke}. In the crescent images, the shadow corresponds to the dark region in the middle of the image. It is most easily seen with baselines oriented across the top and bottom of the crescent. We can make predictions for the appearance of the shadow in future observations by interpolating the best fitting crescent to the uv-plane locations of future baselines. The results for Sgr A* and M87 are shown in Figure \ref{interp}. In M87 especially, the shadow is visible as the local minima in the visibility amplitude. In Sgr A*, the orientation predicted here is $\simeq 20^\circ$ away from ideal, and additional baselines and/or higher sensitivity will be required to detect the shadow if the crescent model is correct.

The crescent model is motivated by the appearance of theoretical images of black hole accretion flows. To check whether it can faithfully reproduce these images, we fit the crescent to sample images from relativistic MHD simulations in exactly the same way as we fit the mm-VLBI data. We arbitrarily pick $5$, $10$ and $20\%$ error bars, and all give similar results. Sample simulated images and the corresponding best fitting crescent models are shown in Figure \ref{simulations}. The images are a standard black hole accretion flow where the angular momentum axis of the infalling gas aligns with the spin axis of the black hole \citep[left,][]{dexteretal2010}; a model where those axes are misaligned by $15^\circ$ \citep[centre,][]{dexterfragile2012}; and an image of the base of an ultra-relativistic jet \citep[right,][]{dexteretal2012}. The best fitting crescents to the simulation images capture the shape and orientations of the simulated images, which confirms that our model fits both observational data as well as theoretical images of accretion discs and even a jet base. Because it can fit both the mm-VLBI data and many theoretical images, the crescent model can be used as a model-independent mediator between the two. Crescent model parameters can be estimated from theoretical images as a crude assessment of their quality of fit to the mm-VLBI data, and best fitting crescent images can be used in the place of simulated images, i.e. to determine when future observations will be able to distinguish between theoretical models. Note that in general we only expect this model to apply if the emission arises from the inner few Schwarzschild radii, where relativistic effects dominate. For example, jet images with emission concentrated farther from the black hole do not have crescent morphologies \citep{broderickloeb2009}.

The crescent model used here is simplistic, especially in the choice of a constant surface brightness. A variety of more complicated options can also be tried, at the expense of additional free parameters. This is necessary, for example, to achieve high quality fits to various simulated images rather than the qualitative examples shown in Figure \ref{simulations}. Several models along these lines are being explored at present (L. Benkevitch et al., in prep.). In addition, the best fitting crescent models have $\tau \approx 0$ (Figure \ref{contourplot}). Allowing $\tau < 0$ would lead to unphysical negative image intensities, but this could be fixed by manually setting those pixel intensities to zero. This would lead to a wider range of possible crescent models, but their Fourier Transform cannot be calculated analytically. 

\section*{acknowledgements}
A.B.K. was supported in part by the Student Opportunity Fund and the ASUC Academic Opportunity Fund at UC Berkeley. We thank S.  Doeleman and L. Benkevitch for useful discussions about this work. 

\footnotesize{
\bibliographystyle{mn2e}

\begin{thebibliography}{31}
\expandafter\ifx\csname natexlab\endcsname\relax\def\natexlab#1{#1}\fi

\bibitem[{{Bardeen}(1973)}]{bardeen1973}
{Bardeen} J.~M., 1973, in Black holes (Les astres occlus), {DeWitt} B.~S.,
  {DeWitt} C., eds., New York: Gordon and Breach, p. 215

\bibitem[{{Bower} {et~al}\mbox{.}(2006){Bower}, {Goss}, {Falcke}, {Backer}, \&
  {Lithwick}}]{bower2006}
{Bower} G.~C., {Goss} W.~M., {Falcke} H., {Backer} D.~C., {Lithwick} Y., 2006,
  \apjl, 648, L127

\bibitem[{{Broderick} {et~al}\mbox{.}(2009){Broderick}, {Fish}, {Doeleman}, \&
  {Loeb}}]{broderick2009}
{Broderick} A.~E., {Fish} V.~L., {Doeleman} S.~S., {Loeb} A., 2009, \apj, 697,
  45

\bibitem[{{Broderick} {et~al}\mbox{.}(2011){Broderick}, {Fish}, {Doeleman}, \&
  {Loeb}}]{brodericketal2011}
---, 2011, \apj, 735, 110

\bibitem[{{Broderick} \& {Loeb}(2009)}]{broderickloeb2009}
{Broderick} A.~E., {Loeb} A., 2009, \apj, 697, 1164

\bibitem[{{Bromley}, {Melia} \& {Liu}(2001){Bromley}, {Melia}, \&
  {Liu}}]{bromley2001}
{Bromley} B.~C., {Melia} F., {Liu} S., 2001, \apjl, 555, L83

\bibitem[{{Cuadra} {et~al}\mbox{.}(2006){Cuadra}, {Nayakshin}, {Springel}, \&
  {Di Matteo}}]{cuadraetal2006}
{Cuadra} J., {Nayakshin} S., {Springel} V., {Di Matteo} T., 2006, \mnras, 366,
  358

\bibitem[{{Dexter}, {Agol} \& {Fragile}(2009){Dexter}, {Agol}, \&
  {Fragile}}]{dexter2009}
{Dexter} J., {Agol} E., {Fragile} P.~C., 2009, \apjl, 703, L142

\bibitem[{{Dexter} {et~al}\mbox{.}(2010){Dexter}, {Agol}, {Fragile}, \&
  {McKinney}}]{dexteretal2010}
{Dexter} J., {Agol} E., {Fragile} P.~C., {McKinney} J.~C., 2010, \apj, 717,
  1092

\bibitem[{{Dexter} \& {Fragile}(2012)}]{dexterfragile2012}
{Dexter} J., {Fragile} P.~C., 2012, arXiv:1204.4454

\bibitem[{{Dexter}, {McKinney} \& {Agol}(2012){Dexter}, {McKinney}, \&
  {Agol}}]{dexteretal2012}
{Dexter} J., {McKinney} J.~C., {Agol} E., 2012, \mnras, 421, 1517

\bibitem[{Doeleman {et~al}\mbox{.}(2012)Doeleman, Fish, Schenck, Beaudoin,
  Blundell, Bower, Broderick, Chamberlin, Freund, Friberg, Gurwell, Ho, Honma,
  Inoue, Krichbaum, Lamb, Loeb, Lonsdale, Marrone, Moran, Oyama, Plambeck,
  Primiani, Rogers, Smythe, SooHoo, Strittmatter, Tilanus, Titus, Weintroub,
  Wright, Young, \& Ziurys}]{doelemanetal2012}
Doeleman S.~S. {et~al.}, 2012, Science

\bibitem[{{Doeleman} {et~al}\mbox{.}(2008){Doeleman}, {Weintroub}, {Rogers},
  {Plambeck}, {Freund}, {Tilanus}, {Friberg}, {Ziurys}, {Moran}, {Corey},
  {Young}, {Smythe}, {Titus}, {Marrone}, {Cappallo}, {Bock}, {Bower},
  {Chamberlin}, {Davis}, {Krichbaum}, {Lamb}, {Maness}, {Niell}, {Roy},
  {Strittmatter}, {Werthimer}, {Whitney}, \& {Woody}}]{doeleman2008}
{Doeleman} S.~S. {et~al.}, 2008, \nat, 455, 78

\bibitem[{{Falcke}, {Melia} \& {Agol}(2000){Falcke}, {Melia}, \&
  {Agol}}]{falcke}
{Falcke} H., {Melia} F., {Agol} E., 2000, \apjl, 528, L13

\bibitem[{{Fish} {et~al}\mbox{.}(2009){Fish}, {Broderick}, {Doeleman}, \&
  {Loeb}}]{fish2009}
{Fish} V.~L., {Broderick} A.~E., {Doeleman} S.~S., {Loeb} A., 2009, \apjl, 692,
  L14

\bibitem[{{Fish} {et~al}\mbox{.}(2011){Fish}, {Doeleman}, {Beaudoin},
  {Blundell}, {Bolin}, {Bower}, {Chamberlin}, {Freund}, {Friberg}, {Gurwell},
  {Honma}, {Inoue}, {Krichbaum}, {Lamb}, {Marrone}, {Moran}, {Oyama},
  {Plambeck}, {Primiani}, {Rogers}, {Smythe}, {SooHoo}, {Strittmatter},
  {Tilanus}, {Titus}, {Weintroub}, {Wright}, {Woody}, {Young}, \&
  {Ziurys}}]{fishetal2011}
{Fish} V.~L. {et~al.}, 2011, \apjl, 727, L36

\bibitem[{{Foreman-Mackey} {et~al}\mbox{.}(2012){Foreman-Mackey}, {Hogg},
  {Lang}, \& {Goodman}}]{emcee}
{Foreman-Mackey} D., {Hogg} D.~W., {Lang} D., {Goodman} J., 2012, ArXiv
  e-prints

\bibitem[{{Fragile} {et~al}\mbox{.}(2007){Fragile}, {Blaes}, {Anninos}, \&
  {Salmonson}}]{fragile2007}
{Fragile} P.~C., {Blaes} O.~M., {Anninos} P., {Salmonson} J.~D., 2007, \apj,
  668, 417

\bibitem[{{Gebhardt} {et~al}\mbox{.}(2011){Gebhardt}, {Adams}, {Richstone},
  {Lauer}, {Faber}, {G{\"u}ltekin}, {Murphy}, \& {Tremaine}}]{gebhardtetal2011}
{Gebhardt} K., {Adams} J., {Richstone} D., {Lauer} T.~R., {Faber} S.~M.,
  {G{\"u}ltekin} K., {Murphy} J., {Tremaine} S., 2011, \apj, 729, 119

\bibitem[{{Gillessen} {et~al}\mbox{.}(2009){Gillessen}, {Eisenhauer}, {Fritz},
  {Bartko}, {Dodds-Eden}, {Pfuhl}, {Ott}, \& {Genzel}}]{gillessen2009}
{Gillessen} S., {Eisenhauer} F., {Fritz} T.~K., {Bartko} H., {Dodds-Eden} K.,
  {Pfuhl} O., {Ott} T., {Genzel} R., 2009, \apjl, 707, L114

\bibitem[{{Goodman} \& {Weare}(2010)}]{goodmanweare}
{Goodman} J., {Weare} J., 2010, Comm. App. Math. Comp. Sci., 5, 65

\bibitem[{{Huang} {et~al}\mbox{.}(2009){Huang}, {Liu}, {Shen}, {Yuan}, {Cai},
  {Li}, \& {Fryer}}]{huang2009}
{Huang} L., {Liu} S., {Shen} Z., {Yuan} Y., {Cai} M.~J., {Li} H., {Fryer}
  C.~L., 2009, \apj, 703, 557

\bibitem[{{Marrone}(2006)}]{marronephd}
{Marrone} D.~P., 2006, PhD thesis, AA(Harvard University)

\bibitem[{{McKinney}, {Tchekhovskoy} \& {Blandford}(2012){McKinney},
  {Tchekhovskoy}, \& {Blandford}}]{mckinneyetal2012}
{McKinney} J.~C., {Tchekhovskoy} A., {Blandford} R.~D., 2012, \mnras, 423, 3083

\bibitem[{{Mo{\'s}cibrodzka} {et~al}\mbox{.}(2009){Mo{\'s}cibrodzka}, {Gammie},
  {Dolence}, {Shiokawa}, \& {Leung}}]{moscibrodzka2009}
{Mo{\'s}cibrodzka} M., {Gammie} C.~F., {Dolence} J.~C., {Shiokawa} H., {Leung}
  P.~K., 2009, \apj, 706, 497

\bibitem[{{Noble} {et~al}\mbox{.}(2007){Noble}, {Leung}, {Gammie}, \&
  {Book}}]{noble2007}
{Noble} S.~C., {Leung} P.~K., {Gammie} C.~F., {Book} L.~G., 2007, Class. and
  Quant. Gravity, 24, 259

\bibitem[{{Pang} {et~al}\mbox{.}(2011){Pang}, {Pen}, {Matzner}, {Green}, \&
  {Liebend{\"o}rfer}}]{pangetal2011}
{Pang} B., {Pen} U.-L., {Matzner} C.~D., {Green} S.~R., {Liebend{\"o}rfer} M.,
  2011, \mnras, 415, 1228

\bibitem[{{Shcherbakov}, {Penna} \& {McKinney}(2012){Shcherbakov}, {Penna}, \&
  {McKinney}}]{shcherbakovetal2012}
{Shcherbakov} R.~V., {Penna} R.~F., {McKinney} J.~C., 2012, \apj, 755, 133

\bibitem[{{Straub} {et~al}\mbox{.}(2012){Straub}, {Vincent}, {Abramowicz},
  {Gourgoulhon}, \& {Paumard}}]{straubetal2012}
{Straub} O., {Vincent} F.~H., {Abramowicz} M.~A., {Gourgoulhon} E., {Paumard}
  T., 2012, \aap, 543, A83

\bibitem[{{Yuan}, {Quataert} \& {Narayan}(2003){Yuan}, {Quataert}, \&
  {Narayan}}]{yuanquataert2003}
{Yuan} F., {Quataert} E., {Narayan} R., 2003, \apj, 598, 301

\bibitem[{{Yuan} {et~al}\mbox{.}(2009){Yuan}, {Cao}, {Huang}, \&
  {Shen}}]{yuan2009}
{Yuan} Y.-F., {Cao} X., {Huang} L., {Shen} Z.-Q., 2009, \apj, 699, 722

\end{thebibliography}

\label{lastpage}

\end{document}